\documentclass{llncs}
\usepackage{amsmath,amssymb}
\usepackage[T1]{fontenc}

\newcommand{\eq}[1]{Eq.~(\ref{#1})}

\newcommand{\eqs}[2]{Eqs.~(\ref{#1}) and (\ref{#2})}

\newcommand{\sref}[1]{Sec.~\ref{#1}}

\newcommand{\dd}{\mathrm{d}}
\newcommand{\fand}{\text{ and }}

\newcommand{\nn}{\nonumber}

\newcommand{\fn}{\footnote}

\begin{document}

\title{How the spherical modes of gravitational waves can be detected despite only seeing one ray}
\author{Alejandro Torres-Orjuela\inst{1}}
\institute{Department of Physics, The University of Hong Kong, Pokfulam Road, Hong Kong\\
\email{atorreso@hku.hk}
}

\maketitle              

\begin{abstract}

The spherical modes of gravitational waves (GWs) have become a major focus of recent detection campaigns due to the additional information they can provide about different properties of the source. However, GW detection is restricted to only detecting one ray and hence it is not obvious how we can extract information about angular properties. In this paper, we introduce a new gauge that makes visible GW detection does not only contain information on the second time derivative but also on the angular derivatives of the GW. In particular, we show that the angular derivatives are of the same order as the time derivatives of the wave thus allowing us to constrain the spherical modes. To further illustrate the detection of the spherical modes, we discuss how the evolution of the orbit of the source and thus the phase of the wave depends on them.

\end{abstract}

\section{Introduction}\label{sec:int}

Since the first detection by LIGO in 2015~\cite{ligo_virgo_2016a} gravitational wave (GW) detection is an increasingly growing field. Up to date, we do not only have detected almost 100 signals from GW sources but also the information we can extract from a signal has increased~\cite{GWTC1,GWTC2,GWTC3}. A particularly interesting part of this `new information' we can get from detection is the spherical modes of GWs, which allow us to obtain information that otherwise would remain latent~\cite{ligo_virgo_2020b,ligo_virgo_2020a,GWTC3}. Some prominent examples are breaking the degeneracy between inclination and distance of the source, detecting the gravitational kicks induced during merger, or detecting a constant center-of-mass velocity of the source~\cite{calderon-bustillo_clark_2018,ligo_virgo_2020b,ligo_virgo_2020a,torres-orjuela_amaro-seoane_2021,torres-orjuela_chen_2021}.

A major contribution to the detection of the spherical modes has to be accredited to recent efforts in developing waveform models that contain information about the subdominant spherical modes~\cite{cotesta_buonanno_2018,cotesta_marsat_2020,khan_ohme_2020,london_khan_2018,ossokine_buonanno_2020,pan_buonanno_2014,field_galley_2014}. Having waveform models containing the spherical modes, matched filtering techniques can be used to extract information about them when detected~\cite{li_2013,sathyaprakash_schutz_2009}. Nonetheless, matched filtering with sophisticated waveform models only allows us to extract the information and is no answer to how the information is detected in the first place.

Let us illustrate this problem in more detail for the spherical modes of GWs. If we have a signal containing information about the spherical modes then matched filtering can tell us that the information is there. But how does the information about the spherical modes get into the signal? Obviously, the answer is that we get the signal from GW detectors -- be them laser interferometry detectors like LIGO, Virgo, KAGRA, TianQin, or LISA~\cite{virgo_2012,lisa_2017,kagra_2019,ligo_2015,tianqin_2016,torres-orjuela_huang_2023} or atom interferometry detectors like AION, MIGA, ZAIGA, MAGIS, or AEDGE~\cite{aion_2020,miga_2018,zaiga_2020,magis_2021,aedge_2020,abend_allard_2023}. However, this is only pushing the question to another level. Because we have to ask now, how do these detectors see the spherical modes of GWs? In the end, spherical modes are related to the angular properties of the source but we only observe the source from one ray thus looking like we should not have any information about spherical properties. Even assuming we see different rays at different detectors, measuring this difference would require much better accuracy than current detectors have. Considering a typical GW source with a strain of $10^{-22}$ at a distance of several $\rm Mpc$ and assuming LIGO and KAGRA are separated by the diameter of the earth, we would require a detection accuracy of $10^{-38}$ to see an angular difference between the incoming rays, which is over 15 orders below what current detectors can achieve~\cite{GWTC3}.

In this paper, we discuss how the spherical modes of GWs can be detected despite only seeing one ray. We structure the discussion as follows. In \sref{sec:wett}, we introduce the equations determining GWs and discuss how the standard traceless-transverse (TT) gauge is established. Afterward, we review in \sref{sec:gwd} the geodesic deviation as the underlying idea of GW detection and show that using the TT gauge spherical modes seem to not be determined. In \sref{sec:sg}, we first discuss briefly the spherical decomposition of GWs and then introduce a new gauge, which we call spherical gauge. We repeat the calculation of the geodesic deviation using the spherical gauge in \sref{sec:gdsg} and show how the angular derivatives of a GW affect a detector. In \sref{sec:phase}, we discuss how the spherical modes affect the phase of the GW to illustrate how their detection is performed. We summarize our results in \sref{sec:res}.

Through the paper, we use geometrical units ($G=c=1$), Einstein's sum convention to sum over indices appearing twice, and Greek indices run over all coordinates while Latin indices only run over the spatial coordinates ($\mu,\nu,\rho,... = 0,1,2,3$ and $i,j,k,... = 1,2,3$).

\section{Wave equation and traceless-transverse gauge}\label{sec:wett}

GWs were first established as a vacuum solution to Einstein's field equations in the weak field limit~\cite{einstein_1916}. In this limit, the metric can be written as
\begin{equation}\label{eq:lie}
    g_{\mu\nu} = \eta_{\mu\nu} + h_{\mu\nu},
\end{equation}
where $\eta_{\mu\nu}$ is the Minkowsky metric and $h_{\mu\nu}$ is the perturbation or GW which is much smaller than 1. The wave is then described by the field or wave equation
\begin{equation}\label{eq:we}
    \Box\bar{h}_{\mu\nu} = 0
\end{equation}
together with the harmonic gauge (also called Lorentz or de Donder gauge)
\begin{equation}\label{eq:hg}
    \partial_\mu \bar{h}^\mu_{~\nu} = 0,
\end{equation}
where $\Box$ denotes the d'Alembert operator, $\bar{h}_{\mu\nu} := h_{\mu\nu} - (\eta_{\mu\nu}/2)h$ is the ``trace-reversed'' wave and $h := \eta^{\mu\nu}h_{\mu\nu}$ is the trace of $h_{\mu\nu}$~\cite{carroll_1997}.

In General Relativity GWs have two degrees of freedom $h_+$ and $h_\times$~\cite{maggiore_2008}, but the requirement for the wave $h_{\mu\nu}$ to be symmetric and imposing the harmonic gauge in \eq{eq:hg} only constrains the solution to six degrees of freedom. Therefore, we are free to perform a coordinate transformation $x^\mu \to x^\mu + \alpha^\mu$, with $\alpha^\mu$ of order $h_{\mu\nu}$ and fulfilling $\Box\alpha^\mu = 0$, so that the trace-reversed wave becomes
\begin{equation}\label{eq:cot}
    \bar{h}_{\mu\nu} \to \bar{h}_{\mu\nu} - \partial_\mu\alpha_\nu - \partial_\nu\alpha_\mu + \eta_{\mu\nu}\partial_\rho\alpha^\rho.
\end{equation}
Note that this transformation is consistent with the harmonic gauge and thus indeed constrains four more degrees of freedom.

To finally constrain the gauge we have to impose four conditions on $\alpha^\mu$. Here we review the standard case of imposing the TT gauge~\cite{maggiore_2008}. We first demand the wave to be trace-free
\begin{equation}\label{eq:tf}
    \bar{h} = 0
\end{equation}
from which follows that the trace-reversed wave and the wave are equal, i.e., $\bar{h}_{\mu\nu} = h_{\mu\nu}$. The second set of constraints we impose is
\begin{equation}\label{eq:vc}
    h_{0i} = 0.
\end{equation}
Imposing this property together with the harmonic gauge gives us $\partial_0 h^0_{~0} = 0$. This implies that $h_{00}$ is constant in time which for a GW is equivalent to saying it vanishes. Therefore, we get
\begin{equation}\label{eq:ntc}
    h_{0\mu} = 0.
\end{equation}

The properties of the wave after imposing the TT gauge are that it is trace-free, making it equal to the trace-reversed wave, and that all components along the time coordinate vanish. However, this does not mean that the wave takes the familiar form of only having components in the plane perpendicular to the wave vector. This particularly simple form is derived for the case of plane waves~\cite{carroll_1997}.

\section{Gravitational wave detection}\label{sec:gwd}

The underlying idea for the detection of GWs using interferometers is that of geodesic deviation~\cite{virgo_2012,lisa_2017,kagra_2019,ligo_2015,tianqin_2016}. In this case, the motion of two particles moving along close geodesics is considered to describe the effect of a gravitational field by describing how the particles deviate from each other. This deviation is described by~\cite{maggiore_2008}
\begin{equation}\label{eq:gede}
    \frac{\dd^2\xi^\alpha}{\dd\tau^2} + R^\alpha_{~\beta\gamma\delta}\frac{\dd x^\beta}{\dd\tau}\xi^\gamma\frac{\dd x^\delta}{\dd\tau} = 0,
\end{equation}
where $\xi^\alpha$ is the deviation vector connecting the two particles, $R^\alpha_{~\beta\gamma\delta}$ is the Riemann tensor~\cite{misner_thorne_1973}, $x^\beta$ denotes the geodesic and $\tau$ is the proper time along the geodesic.

We simplify \eq{eq:gede} by considering a detector at rest. In this case the derivatives of the geodesic reduce to the time vector and the proper time is equal to coordinate time~\cite{torres-orjuela_chen_2019}. Thus, we get
\begin{equation}\label{eq:gdr}
    \partial_0\partial_0\xi^\alpha + R^\alpha_{~0\gamma0}\xi^\gamma = 0.
\end{equation}
Last we can use that for GWs the Riemann tensor takes the form~\cite{maggiore_2008}
\begin{equation}\label{eq:riem}
    R^\alpha_{~\beta\gamma\delta} = \frac12\eta^{\alpha\rho}(\partial_\beta\partial_\gamma h_{\delta\rho} + \partial_\delta\partial_\rho h_{\beta\gamma} - \partial_\gamma\partial_\rho h_{\beta\delta} - \partial_\beta\partial_\delta h_{\gamma\rho})
\end{equation}
to get
\begin{equation}\label{eq:gwgd}
    \partial_0\partial_0\xi^i - \frac12\delta^{ij}(\partial_0\partial_k h_{0j} + \partial_0\partial_j h_{0k} - \partial_j\partial_k h_{00} - \partial_0\partial_0 h_{jk})\xi^k = 0,
\end{equation}
where $\delta_{ab}$ denotes the Kronecker delta and we ignored $\xi^0$ because GWs do not affect the time component of a 4-vector while the spatial components of the same vector are not affected by its time component when interacting with GWs.

If we now consider a GW in the usual TT gauge where $h_{0\mu} = 0$, \eq{eq:gwgd} reduces to
\begin{equation}\label{eq:gdtt}
    \partial_0\partial_0\xi^i + \frac12\delta^{ij}\partial_0\partial_0h_{jk}\xi^k = 0.
\end{equation}
Thus we recover the classical statement that the arm of a detector is stretched and squeezed in lockstep with the GW. However, we do not see how the spherical modes of GWs could be detected. The deviation of the geodesic only depends on the time derivative of the wave and because we only see one ray this means there should be no information about the angular properties. Having said this, it is important to note that this is a description of the wave in only one particular gauge. Using another gauge the other information contained by the wave will become more obvious.

\section{Spherical gauge}\label{sec:sg}

We want to introduce a new gauge for GWs which we call \textit{spherical gauge}. This gauge will allow us to show how the spherical modes of the wave can be detected. However, before introducing the spherical gauge we give a brief overview of the decomposition of GWs in spherical modes.

The two polarizations of GWs $h_+$ and $h_\times$ can be combined to define the so-called complex amplitude~\cite{ruiz_alcubierre_2008}
\begin{equation}\label{eq:camp}
    H := h_+ - ih_\times,
\end{equation}
where $i$ is the imaginary unit. This complex amplitude can then be decomposed using spin=$-2$ spherical harmonics, $_{-2}Y^{\ell,m}(\theta,\phi)$, to obtain~\cite{goldberg_macfarlane_1967,ruiz_alcubierre_2008}
\begin{equation}\label{eq:moex}
    H(t,r,\theta,\phi) = \sum_{\ell=2}^\infty\sum_{m=-\ell}^\ell H^{\ell,m}(t,r)_{-2}Y^{\ell,m}(\theta,\phi),
\end{equation}
where
\begin{equation}\label{eq:mode}
    H^{\ell,m}(t,r) := \int\dd\Omega H(t,r,\theta,\phi)_{-2}\bar{Y}^{\ell,m}(\theta,\phi)
\end{equation}
are the GW spherical modes and $_{-2}\bar{Y}^{\ell,m}(\theta,\phi)$ means the complex conjugate of the spin=$-2$ spherical harmonics. The spherical modes then contain the information about the evolution of the wave in time and its dependence on the distance, while the $_{-2}Y^{\ell,m}(\theta,\phi)$ contain the information about the angular properties of the wave.

The spin=$-2$ spherical harmonics represent an orthonormal base for functions of the angular coordinates $(\theta,\phi)$
\begin{equation}\label{eq:onb}
    \int\dd\Omega _{-2}Y^{\ell,m}(\theta,\phi)_{-2}\bar{Y}^{\ell',m'}(\theta,\phi) = \delta_{\ell\ell'}\delta_{mm'}.
\end{equation}
Moreover, they have relatively simple derivatives along the angular coordinates
\begin{align}\label{eq:shdet}
    \nn \partial_\theta\,_{-2}Y^{\ell,m}(\theta,\phi) =& \frac12\bigg(\sqrt{(\ell-m)(\ell+ m+1)}_{-2}Y^{\ell,m+1}(\theta,\phi)e^{-i\phi} \\
    &- \sqrt{(\ell+m)(\ell-m+1)}_{-2}Y^{\ell,m-1}(\theta,\phi)e^{+i\phi}\bigg), \\ \label{eq:shdep}
    \partial_\phi\,_{-2}Y^{\ell,m}(\theta,\phi) =& im_{-2}Y^{\ell,m}(\theta,\phi).
\end{align}

\subsection{The spherical gauge}\label{ssec:sg}

Let us now introduce the spherical gauge. A GW has to fulfill \eqs{eq:we}{eq:hg}, while we are still free to use a coordinate transformation as in \eq{eq:cot}. By this additional coordinate transformation, we constrain four more degrees of freedom so that the wave only has the two polarizations $h_+$ and $h_\times$~\cite{maggiore_2008}.

From the coordinate transformation, we are free to impose four conditions on the wave. If the trace of a matrix vanishes in one coordinate system it also vanishes in all other coordinate systems. Therefore, we (have to) keep the condition that the trace of the wave vanishes, $\bar{h} = 0$. This condition again guarantees that the trace-reversed wave and the wave are equal. The second set of conditions we impose is that
\begin{equation}\label{eq:sgc}
    h_{ri} = 0.
\end{equation}
This condition is similar to the condition of the TT gauge (cf. \eq{eq:vc}) but now the radial components instead of the time components vanish.

Analogous to the case of the TT gauge, if the $h_{ri}$ vanish $h_{0r}$ also has to vanish. Therefore, we get that in spherical gauge the wave is trace-free and the radial components are all equal to zero
\begin{equation}\label{eq:psg}
    h = 0 \fand h_{r\mu} = 0.
\end{equation}
Note that the spherical gauge and the TT gauge only differ by a constant rotation of the time coordinate and the radial coordinate. Therefore, the spherical coordinates $\theta$ and $\phi$, which we want to study, are not affected by the transformation and can be interpreted in the usual way we know from the TT gauge. Before we move on to discuss the detection of GWs expressed in spherical gauge, we discuss the properties of the wave.

\subsection{Time components in spherical gauge}\label{ssec:tcsg}

We know a GW has only two degrees of freedom $h_+$ and $h_\times$~\cite{maggiore_2008}. Therefore, we can express any component of the wave as
\begin{equation}\label{eq:wlc}
    h_{\mu\nu} = p_{\mu\nu}h_+ + c_{\mu\nu}h_\times,
\end{equation}
where $p_{\mu\nu}$ and $c_{\mu\nu}$ are real constant numbers.

Let us assume all $h_{0\mu}$ vanish. In this case we have $p_{0\mu} = c_{0\mu} = 0$ and from the trace free condition we get $p_{\phi\phi} = -p_{\theta\theta}$ and $c_{\phi\phi} = -c_{\theta\theta}$. Using this information together with the harmonic gauge in \eq{eq:hg} for the $\theta$ and $\phi$ coordinates gives us
\begin{align}\label{eq:hgt}
    0 =& p_{\theta\theta}\partial_\theta h_+ + c_{\theta\theta}\partial_\theta h_\times + p_{\theta\phi}\partial_\phi h_+ + c_{\theta\phi}\partial_\phi h_\times, \\ \label{eq:hgp}
    0 =& p_{\theta\phi}\partial_\theta h_+ + c_{\theta\phi}\partial_\theta h_\times - p_{\theta\theta}\partial_\phi h_+ - c_{\theta\theta}\partial_\phi h_\times.
\end{align}
Summing the two equations we then get
\begin{equation}\label{eq:hgs}
    0 = (p_{\theta\phi} + p_{\theta\theta})\partial_\theta h_+ + (c_{\theta\phi} + c_{\theta\theta})\partial_\theta h_\times + (p_{\theta\phi} - p_{\theta\theta})\partial_\phi h_+ + (c_{\theta\phi} - c_{\theta\theta})\partial_\phi h_\times.
\end{equation}

Using that $h_+ = \Re[H]$ and $h_\times = -i\Im[H]$ (cf. \eq{eq:camp}) and the partial derivatives of the spin=$-2$ harmonics in \eqs{eq:shdet}{eq:shdep} we can see that the derivatives of the polarizations all differ by more than just a constant factor. Therefore, the $\partial_\theta h_+$, $\partial_\phi h_+$, $\partial_\theta h_\times$ and $\partial_\phi h_\times$ are linearly independent and \eq{eq:hgs} is only fulfilled if all coefficients vanish independently. However, this is only possible for
\begin{equation}
    p_{\theta\theta} = p_{\theta\phi} = c_{\theta\theta} = c_{\theta\phi} = 0,
\end{equation}
thus implying that the whole wave vanishes.

In a similar way it can be shown that the $h_{0i}$ have to vanish if $h_{00} = 0$, thus again implying that the whole wave would vanish. Analyzing again the different cases where only one of the $h_{0\mu}$ is equal to zero, we find that $h_{00}$ never vanishes and that only one of $h_{0\theta}$ and $h_{0\phi}$ can vanish at the same time ($h_{0r} = 0$ from the spherical gauge condition).

\subsection{Derivatives of the wave in spherical gauge}~\label{ssec:dwsg}

Any system that emits GWs has a different number of dynamical fields that travel away from the source~\cite{misner_thorne_1973}. However, only those fields that decrease as $1/r$, where $r$ is the distance from the source\fn{Note that the $r$ we refer here to is the distance between the source and the observer which for the spherical gauge can differ from the radial coordinate $r$ due to the rotation applied to the time coordinate and the radial coordinate.}, are called GWs and can be detected by a distant observer~\cite{maggiore_2008}. Therefore, for detection, it is important to understand which components of the GW decrease as $1/r$.

We start noticing that if the component of a GW $h_{\mu\nu}$ decreases as $1/r$, its first and second time derivatives also decrease as $1/r$ to the leading order~\cite{maggiore_2008}, i.e.,
\begin{equation}\label{eq:tdlo}
    \partial_0h_{\mu\nu}, \partial_0\partial_0h_{\mu\nu} \propto \frac1r.
\end{equation}
However, for the spatial derivatives, this is not necessarily true, and in particular, the derivatives along the angular coordinates $\theta$ and $\phi$ often decrease at higher orders.

We analyze how the angular derivatives of a GW in spherical gauge decrease to judge if they can be detected. Using the harmonic gauge in \eq{eq:hg}, we get
\begin{equation}\label{eq:fde}
    \partial_i h_{ij} = \partial_0 h_{0j}.
\end{equation}
Deriving the previous equation along $j$ and applying again the harmonic gauge condition, we then find
\begin{equation}\label{eq:sde}
    \partial_i\partial_j h_{ij} = \partial_0\partial_0 h_{00}.
\end{equation}

From the analysis in the previous section, we know that $h_{00}$ cannot vanish, which for GWs is analogous to saying it decreases as $1/r$. Now because its second time derivative also decreases as $1/r$ and from the spherical gauge condition in \eq{eq:sgc} we know the radial components vanish, we see from \eq{eq:fde} that to leading order
\begin{equation}\label{eq:dede}
    \partial_\theta\partial_\theta h_{\theta\theta}, \partial_\phi\partial_\phi h_{\phi\phi}, \partial_\theta\partial_\phi h_{\theta\phi} \propto \frac1r.
\end{equation}

Therefore, we get that in spherical gauge the second derivatives of the wave along the angular coordinates decrease as $1/r$, at least for some of the components. That means that they, in principle, can be detected in the same way as other properties of the wave. Now it only remains to analyse to what extent a detector is sensitive to these derivatives.

\section{Geodesic deviation in spherical gauge}\label{sec:gdsg}

We reconsider the geodesic deviation induced by a GW but now using the spherical gauge. Applying the spherical gauge none of the terms in \eq{eq:gwgd} vanish and then using the harmonic gauge in \eq{eq:hg}, we find for the geodesic deviation
\begin{equation}\label{eq:gdsg}
    \partial_0\partial_0\xi^i - \frac12\delta^{ij}(\partial_i\partial_k h_{ij} + \partial_i\partial_j h_{ik} - \partial_j\partial_k h_{00} - \partial_0\partial_0 h_{jk})\xi^k = 0
\end{equation}

We see from \eq{eq:gdsg} that in spherical gauge the geodesic deviation in principle can depend on the angular derivatives of the wave. We further know from the analysis in \sref{ssec:dwsg} that several of the derivatives along the angular coordinates decrease as $1/r$ and thus are detectable.

Nevertheless, for better comprehension, we consider a two-arm detector where one of the arms lies along the $\theta$ and the other along the $\phi$ coordinate. In this case and only keeping those terms decreasing as $1/r$, \eq{eq:gdsg} reduces to
\begin{align}\label{eq:deth}
    \partial_0\partial_0\xi_\theta - \left(\partial_\theta\partial_\theta h_{\theta\theta} + \partial_\theta\partial_\phi h_{\theta\phi} - \frac12\partial_0\partial_0 h_{\theta\theta}\right)\xi_\theta =& 0, \\ \label{eq:deph}
    \partial_0\partial_0\xi_\phi - \left(\partial_\phi\partial_\phi h_{\phi\phi} + \partial_\theta\partial_\phi h_{\theta\phi} - \frac12\partial_0\partial_0 h_{\phi\phi}\right)\xi_\phi =& 0.
\end{align}
This means that when expressing a GW in spherical gauge, we find that the detection depends on its second angular derivatives. In addition, we see that the detection still depends on the second time derivative of the wave and thus reduces to the case known for the TT gauge (cf. \eq{eq:gdtt}) when ignoring the spherical properties of the wave.

\section{Dependence of the phase on the spherical modes}\label{sec:phase}

We have shown that GWs contain information about the spherical modes even when detecting only one ray. Therefore, we are in principle able to detect them although it is still not obvious how we detect spherical modes in practice, in particular, when the TT gauge is used to interpret detection. To clarify this last point, we discuss in this section how the spherical modes affect the main observable in GW detection -- the phase of the wave~\cite{sathyaprakash_schutz_2009}.

The relation between the phase of a GW and its spherical modes is, in general, non-trivial due to the nonlinearity of General Relativity and how the spherical modes change as the phase evolves. Therefore, we consider the illustrative case of a pair of compact objects far from the merger. In this case, the orbit of the binary can be approximated to be Keplerian, and establishing a relation between the phase and the spherical modes is possible. Before we derive this relation, we point out that this case is illustrative but not unrealistic as GW sources can have a significant number of strong spherical modes far from the merger when they have a high eccentricity or there is a high mass ratio between the two components of the binary~\cite{drasco_hughes_2006,peters_mathews_1963,torres-orjuela_2023}.

For a Keplerian orbit, the total energy of the source and the period of the orbit are, respectively,
\begin{align}\label{eq:kepen}
    E &= \frac{M}{2a}, \\ \label{eq:keppe}
    P &= 2\pi\sqrt{\frac{a^3}{M}}.
\end{align}
Combining these two equations and using that the orbital frequency of a Keplerian orbit is $\omega_o := 2\pi/P$, we get
\begin{equation}\label{eq:orbf}
    \omega_o = \frac{\sqrt{8E^3}}{M}.
\end{equation}

The energy carried away from the source by GWs to infinity can be expressed in terms of the spherical modes~\cite{gerosa_hebert_2018}
\begin{equation}\label{eq:enmode}
    \dot{E} = \frac{1}{16\pi}\sum_{\ell,m}\left|\dot{h}^{\ell,m}\right|^2,
\end{equation}
where a dot indicates a time derivative, $h^{\ell,m} := rH^{\ell,m}$, and we ignored the time derivative of $1/r$ as it decreases as $1/r^2$, $r$ being the distance between the source and the observer.

Taking the time derivative of the orbital frequency in \eq{eq:orbf} and using \eq{eq:enmode}, we obtain
\begin{equation}\label{eq:fremode}
    \dot{\omega}_o = \frac{3\sqrt{8}}{32\pi M}\sqrt{E}\sum_{\ell,m}\left|\dot{h}^{\ell,m}\right|^2.
\end{equation}
The phase of a GW is the integral of the frequency of the wave over time $\Phi_{\rm GW} = \int\omega_{\rm GW}\dd t$ while the frequency of the wave is proportional to the orbital frequency of the source's orbit. Therefore, we get that the phase of the wave is proportional to the time integral of the sum of the absolute value of the spherical modes
\begin{equation}\label{eq:phamode}
    \Phi_{\rm GW} \propto \int\sum_{\ell,m}\left|\dot{h}^{\ell,m}\right|^2\dd t.
\end{equation}

From \eq{eq:phamode}, we see that two sources with different spherical modes will have a different GW phase due to their different evolution in time. At the same time, it becomes obvious why we need a high signal-to-noise ratio to detect the spherical modes as weak sub-dominant modes only change the phase marginally. Moreover, if the difference in the spherical modes of two sources is negligible then we are not able to turn them apart due to a similar evolution of their phase.

\section{Results}\label{sec:res}

The spherical modes of GWs have become a major focus of recent detection campaigns due to the additional information they can provide. Nevertheless, detection is usually restricted to say we have detected spherical modes because of a better result when using matched filtering techniques. Although this may work well in an everyday business it is no explanation to the fundamental question of detecting a physical property. One even could think that by detecting only one ray of a GW, we should not be able to extract any information about the angular properties of the source and hence the spherical modes. At first glance, this picture even seems to be confirmed when considering detection using the usual TT gauge.

In this paper, we introduced a new gauge, which we call spherical gauge. We showed that using the spherical gauge the geodesic deviation and hence detection does not only depend on the second time derivative of the wave -- as for the TT gauge -- but also on the angular derivatives. We further consider the behavior of the angular derivatives to show that they decrease as $1/r$, thus being of the same order as the wave. That the detection also depends on the angular derivatives makes clear why we can detect GW spherical modes. Having information about the value of a function and its derivatives allows us to constrain an equation to a better degree. This is equally true for the spin=$-2$ spherical harmonics which are the basis of spherical mode decomposition. To better illustrate the actual detection of spherical modes, we discuss how spherical modes relate to the evolution of the source's orbit and thus the phase of the GW. We see that using an appropriate gauge the information GW detection contains about the spherical properties of the source becomes evident and we understand why this information can be extracted using matched filtering techniques.

\section*{Acknowledgment}

Many thanks to Xian Chen for numerous challenging and thus extremely helpful discussions on the detection of gravitational wave spherical modes. I further thank Carlos F. Sopuerta for insightful comments about the effect of spherical modes on the phase of GWs. This work was partially supported by the Guangdong Major Project of Basic and Applied Basic Research (Grant No. 2019B030302001).


\bibliographystyle{plain}
\bibliography{alebib.bib}
\end{document}